\documentclass[runningheads]{llncs}
\renewcommand{\thefootnote}{\fnsymbol{footnote}}
\usepackage{graphicx}
\usepackage{comment}
\usepackage{amsmath,amssymb}
\usepackage{color}
\usepackage{url}
\usepackage{hyperref}

\usepackage[utf8]{inputenc} % allow utf-8 input
\usepackage[T1]{fontenc}    % use 8-bit T1 fonts
\usepackage{url}            % simple URL typesetting
\usepackage{booktabs}       % professional-quality tables
\usepackage{amsfonts}       % blackboard math symbols
\usepackage{nicefrac}       % compact symbols for 1/2, etc.
\usepackage{microtype}      % microtypography
\usepackage{lipsum}		% Can be removed after putting your text content
\usepackage{graphicx}
\usepackage{makecell}
\usepackage{svg}
\usepackage{algorithm}
\usepackage{algpseudocode}
\usepackage{amsmath}   % For math symbols like \mathcal
\usepackage{amssymb}   % For symbols like \setminus
\usepackage[numbers]{natbib}
\usepackage{doi}
\usepackage{color}
\usepackage{subcaption,caption,algorithm,algorithmicx,amsmath,enumitem,cleveref}
\usepackage{xr-hyper} % or \usepackage{xr-hyper} if using hyperref
\newif\ifreview
\reviewfalse

\ifreview
	\usepackage{lineno}

	\linenumbers
\fi

\begin{document}

%%%%%%%%%%%%%%%%%%%%% Add submission id, track, and title. %%%%%%%%%%%%%%%%%%%%%

% TODO: Please insert your submission number here
\def\SubNumber{009}

% TODO: Please uncomment the track this paper will be submitted to, comment all other lines
%\def\GCPRTrack{Main Track}
\def\GCPRTrack{Special Track: Pattern recognition in the life and natural sciences}
%\def\GCPRTrack{Special Track: Photogrammetry and remote sensing}
%\def\GCPRTrack{Young Researcher's Forum}
%\def\GCPRTrack{Fast Review Track}

% TODO: Replace with your title
\title{subCellSAM: Zero-Shot (Sub-)Cellular Segmentation for Hit Validation in Drug Discovery}
% You can use \thanks for acknowledgment. Do not add any acknowledgment to the draft 
% version that is used for the review process.  
%\title{Title\thanks{XXX}}

\ifreview
	% ANONYMOUS SUBMISSION FOR REVIEW
	% DO NOT MODIFY these for the draft version that is used for the review process.
	\titlerunning{GCPR 2025 Submission \SubNumber{}. CONFIDENTIAL REVIEW COPY.}
	\authorrunning{GCPR 2025 Submission \SubNumber{}. CONFIDENTIAL REVIEW COPY.}
	\author{GCPR 2025 - \GCPRTrack{}}
	\institute{Paper ID \SubNumber}
\else
	% CAMERA READY SUBMISSION
	%\titlerunning{Abbreviated paper title}
	% If the paper title is too long for the running head, you can set
	% an abbreviated paper title here

	\author{Jacob Hanimann\thefootnote{$^\ast$}\thefootnote{$^\ddagger$} \inst{1} \and
	Daniel Siegismund\thefootnote{$^\ast$} \inst{2} \and Mario Wieser\inst{2} \and Stephan Steigele\inst{2}}
	
	\authorrunning{J. Hanimann et al.}
    \titlerunning{subCellSAM}
	% First names are abbreviated in the running head.
	% If there are more than two authors, 'et al.' is used.
	
	\institute{University of Bern, Bern, Switzerland \and Genedata AG, Basel, Switzerland}
\fi

\maketitle              % typeset the header of the contribution
\def\thefootnote{$\ast$}\footnotetext{These authors contributed equally to this work.}
\def\thefootnote{$\ddagger$}\footnotetext{Work done during an internship at Genedata AG.}
\begin{abstract}
High-throughput screening using automated microscopes is a key driver in biopharma drug discovery, enabling the parallel evaluation of thousands of drug candidates for diseases such as cancer. Traditional image analysis and deep learning approaches have been employed to analyze these complex, large-scale datasets, with cell segmentation serving as a critical step for extracting relevant structures. However, both strategies typically require extensive manual parameter tuning or domain-specific model fine-tuning. We present a novel method that applies a segmentation foundation model in a zero-shot setting (i.e., without fine-tuning), guided by an in-context learning strategy. Our approach employs a three-step process for nuclei, cell, and subcellular segmentation, introducing a self-prompting mechanism that encodes morphological and topological priors using growing masks and strategically placed foreground/background points. We validate our method on both standard cell segmentation benchmarks and industry-relevant hit validation assays, demonstrating that it accurately segments biologically relevant structures without the need for dataset-specific tuning.


\end{abstract}
\section{Introduction}
The development of a new drug is a prolonged and costly process that takes more than ten years with up to four billion dollars of investment \cite{Sertkaya}. Consequently, the discovery of new drug candidates is a key driving factor in the process of developing sophisticated treatment strategies such as cancer immunotherapies \cite{cancerim}. \\
In early drug discovery, high-content screening (HCS) has become a key technology to assess the effects of chemical compounds on cellular systems \cite{ZANELLA2010237}. By integrating automated microscopy with advanced image analysis, HCS enables the collection of high-dimensional data from individual cells, and as it is run on 384-1536 microtiter plates it allows for the parallel testing of tens of thousands of compounds per experiment. This approach provides detailed spatial and temporal insights into cellular responses, supporting the identification of biologically relevant mechanisms. Among the various HCS methodologies, multiplexed imaging assays such as cell painting have been developed to capture diverse morphological features \cite{Bray2016}, though many other assay formats are also employed depending on the biological context and research objectives.\\
A major challenge in HCS assays is the need to analyze both large-volume and biologically complex datasets. Traditionally, such datasets have been analyzed by employing handcrafted features with classical image analysis software such as CellProfiler \cite{carpenter2006cellprofiler} which is highly time-consuming and biased as the analysis pipeline has to be set up manually by a domain expert for each screening dataset \cite{Caicedo2017DataanalysisSF}. Hence, deep learning based approaches have been developed to speed up the process by learning image-based representations without requiring the manual set up of image analysis pipelines\cite{STEIGELE2020812,godinez2017multi,ChannelImportance,pmlr-v172-siegismund22a,9911974,10944004}.
A crucial step in performing this analysis includes the segmentation of cells and their corresponding compartments from high-content images to extract all relevant structures. This step is particularly important as errors will propagate through the analysis and may influence the downstream result calculations and decision processes. While early approaches relied on classical image segmentation techniques \cite{carpenter2006cellprofiler}, more sophisticated methods specifically targeted at cells have since been introduced in recent years \cite{stringer2025cellpose3,van2016deep,israel2023foundationmodelcellsegmentation,tang2024generative}. However, all of the above mentioned methods either require extensive manual parameter tuning (CellProfiler) or specific training or fine-tuning for cell-segmentation tasks.\\
To address this challenge, we propose an approach that leverages a pre-trained segmentation foundation model without any task-specific fine-tuning, guided by an in-context learning strategy. This strategy is designed to embed key morphological and topological priors characteristic of cell images into the prompting process. Specifically, we first perform nuclei segmentation, followed by an iterative segmentation of the cell body using a loop prompting mechanism that enforces these priors and incorporates both positive and negative anchor points, along with low-resolution masks from previous iterations.

Our contributions can be summarized as follows:
\begin{itemize}
    \item We propose a novel in-context learning approach for (sub)cellular segmentation based on iterative self-prompting.
    \item We introduce a prompt sampling strategy that encodes key morphological and topological priors, including cell-to-nucleus relationships and instance separation, to enhance pipeline robustness and generalizability.
    \item We evaluate our method for cell body segmentation across three diverse datasets, where it shows performance competitive with or exceeding specialized methods.
    \item We validate the applicability of our method on two hit-validation datasets relevant to drug discovery, where it outperforms the consensus baseline.
\end{itemize}

% Related Work section
\section{Related Work}
\subsubsection{In-Context Learning}
In contrast to model training approaches, in-context learning aims to solve specific tasks by conditioning on a certain amount of demonstrations without updating the model weights. Depending on the number of demonstrations which are required to solve a certain task, in-context learning can be further divided into sub-groups: Few-shot learning aims to solve a specific task by providing a number of examples to the model. Especially, large-language models have demonstrated to perform well in such tasks \cite{brown}. Moreover, few-shot learning in the context of foundation models has been utilized in various tasks such as object detection \cite{Han_2024_CVPR}. In addition, zero-shot learning aims to solve tasks without providing any demonstrations to the model. Here, large language models have been used to perform zero-shot captioning of images \cite{tewel2021zero}, gaze following \cite{Gupta_2024_CVPR} or on lesion detection \cite{Guo}.

\subsubsection{Vision-Language Models}
In recent years, large language models have been combined with vision models to reason across multiple modalities. CLIP learns visual concepts from natural language supervision to solve vision tasks across multiple domains \cite{clip}. In addition, \cite{flamingo} introduced a vision-language model to solve tasks like visual question-answering in a few-shot fashion. In addition to the previously mentioned tasks, segmentation of objects plays a crucial role in various application areas. Kirillov et al. \cite{Kirillov_2023_ICCV} introduced SAM, a segmentation approach to allow for zero-shot segmentation of new imaging tasks. This approach has subsequently been extended in multiple directions: \cite{NEURIPS2023_5f828e38} introduced improvements to segment objects in higher quality while \cite{zhao2023fastsegment} have proposed a CNN backbone instead of a transformer architecture to speed up the segmentation process. More recently, SAM was extended to segment video sequences \cite{ravi2025sam}.

\subsubsection{Cell-Based Segmentation Models}
Building on their success in natural image analysis, vision-language segmentation models have recently been extended to cell segmentation. \cite{na2024segmentcellsambasedautoprompting} proposed an extension of SAM to enable nuclei segmentation while \cite{seg_foundation_model} introduced a foundation model to segment any kind of microscopy images. Additionally, \cite{stringer2025cellpose3} demonstrated a method for improved cellular segmentation  for images which suffer from noise, blurring or undersampling, a common issue in cell microscopy. \cite{israel2023foundationmodelcellsegmentation} suggested a method for cell segmentation which trains an object detector for cell detection in combination with a prompt engineering approach to generate segmentations. In contrast to the previous approaches, \cite{van2016deep} developed a segmentation method based on convolutional networks for live-cell imaging experiments. These approaches require at least partial retraining of certain parameters.

\section{Setup and Preliminaries}
Consider a dataset $D = \{x_i\}_{i=1}^N$ which consists of $N$ data samples. Here, $x_i \in \mathbb{R}^{C \times W \times H}$ represents a microscopy image with $C$ channels and dimensions $W$ and $H$. More specifically, we base our approach on the Segment Anything Model (SAM) \cite{Kirillov_2023_ICCV}, an image segmentation model which can be guided by prompts consisting of either text, masks, bounding boxes or points in an image. SAM consists of three main components: An image encoder which maps an image $x$ into a latent representation $z$. A flexible prompt encoder to guide the segmentation process depending on the prompt input. In our approach, we employ image mask and point prompts which are mapped via an encoder into a latent representation $p$ with equal size of $z$ which is subsequently combined with $z$ \cite{NEURIPS2020_55053683}.
Finally, a segmentation mask decoder is employed which generates an output mask based on the image embedding $z$ and a set of prompt embeddings $p$. For more detailed information please refer to \cite{Kirillov_2023_ICCV}. 
In the remaining part of this paper, we assume that our method adheres to the following three assumptions:
\noindent\textbf{Assumption 1: Nuclei marker channel} \textit{We assume that each image contains a nuclei marker channel  (e.g., DAPI, Hoechst) as prior information for cell segmentation.}
\noindent\textbf{Assumption 2: Cell shape marker channel} \textit{In order to perform cell segmentation, we require at least one corresponding cell marker channel (e.g., membrane or cytoplasmic stain) which marks the cell boundaries.}
\noindent\textbf{Optional Assumption 3: Subcellular structures channel} \textit{In case we aim to segment subcellular structures, we assume to have a dedicated subcellular marker channel.}

\section{Method}
\label{sec:methods}

\begin{figure}[ht]
  \centering
  % grab page 1 of the PDF and make it span the text width
  \includegraphics[page=1,width=\textwidth]{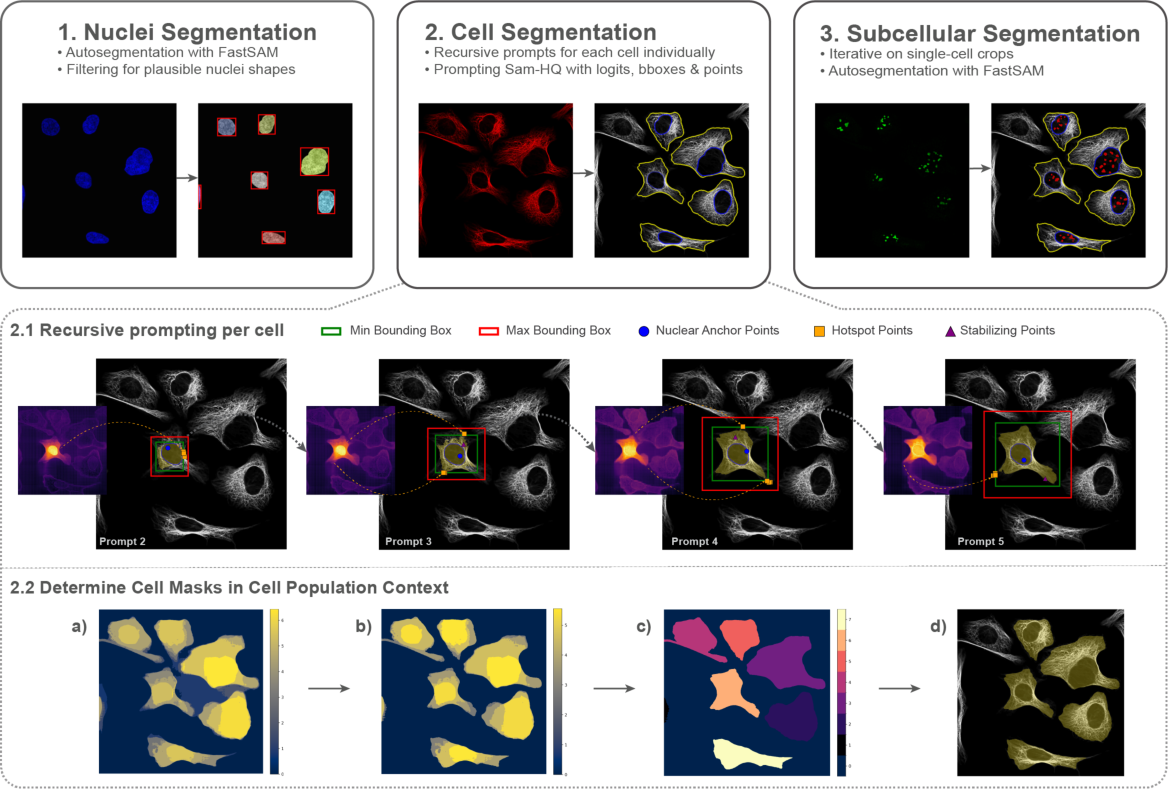}
 \caption{\textbf{subCellSAM Workflow Overview.} Human Protein Atlas \cite{HPA} example: UTP6 (green; nucleolus, small-subunit processome), microtubules (red), and nucleus (blue). Modules: (1) \textit{Nuclei}, (2) \textit{Cell}, and (3) \textit{Subcellular Segmentation}. 
    Panel 2.1: Recursive prompting per cell; iterations refine segmentation. Prompts: \textit{Hotspot Points}; sampled within \textit{Min/Max Bbox} using the mean logits map from the two previous masks as a probability map, \textit{Stabilizing Points}, and \textit{Nucleus Anchor Points}. 
    Panel 2.2: Cell mask integration: low coverage regions are removed (a-b), pixels assigned to cell ID/background (c), resulting in final boundaries (d).}
  \label{overview}
\end{figure}

We propose subCellSAM, an in-context learning approach for zero-shot segmentation of single cells and their subcellular compartments from multi-channel fluorescence microscopy images. A detailed description of the method may be found in Figure \ref{overview} and an algorithmic explanation is illustrated in Algorithm \ref{algo:overall}. More specifically, our method is divided into three distinct parts. First, we perform nuclei segmentation to obtain segmentation masks as a starting point for cell segmentation. Subsequently, we enable an in-context learning strategy to perform cell and subcellular entity segmentation as step two and three, respectively.

\begin{algorithm}[ht]
\caption{Zero-Shot Subcellular Segmentation}
\label{alg:cell_segmentation}
\begin{algorithmic}[1]
\Require Microscopy Image $x \in \mathbb{R}^{c \times w \times h}$, max iterations $I_{max}$
\State $\{y\}, \{y_{\text{center}}\}$ $\gets$ \text{getNucleusMaskandCenter(x)}  \Comment{Refer to Section \ref{promt:nuclei}}
\State cellSeg $\gets$ []
\State subcellSeg $\gets$ []
\For{mask in $\{y\}$}
	\State cell $\gets$ segmentCell(x, mask) \Comment{Refer to Section \ref{sec:cellularseg}}
    \State subcell $\gets$ segmentSubCell(x, mask) \Comment{Refer to Section \ref{sec:subcellularseg}}
    \State cellSeg.append(cell), subcellSeg.append(subcell)    	
\EndFor
\State \textbf{return} cellSeg, subcellSeg
\end{algorithmic}
\label{algo:overall}
\end{algorithm}

\subsection{ Nuclei Segmentation}
\label{promt:nuclei}
Nuclei segmentation is performed by employing segmentation followed by mask filtering. For each microscopy image $x \in \mathbb{R}^{C \times W \times H}$ , we perform nuclei segmentation on the defined nuclei marker channel (see Figure~\ref{overview}, Step 1). To do so, we feed the nuclei channel $nm \in \mathbb{R}^{1 \times W \times H}$ into a pretrained segmentation model $f$ (e.g. FastSAM \cite{zhao2023fastsegment}) in order to obtain an initial list of candidate nuclei masks $\{y\}_{i=1}^m$ where $y$ denotes the mask and $m$ the number of candidate masks per image. Note that $m$ may vary for different images. This process is leveraged by SAM's automatic mask generation module. Subsequently, these candidate masks are filtered to isolate valid nuclei from imaging artifacts, debris, or multi-nucleus conglomerates. This is achieved through statistical outlier detection. For each mask, we compute its area, aspect ratio, and circularity. We then discard any masks whose properties fall significantly outside the population's norm (e.g., more than two standard deviations from the median area), a process that effectively removes implausible objects. The validated nucleus masks are further used as seeds for cell segmentation step. The algorithm is described in the Appendix~\ref{algo:nucl}.

\subsection{Cell Segmentation Guided by Morphological and Topological Priors}
\label{sec:cellularseg}
Based on the predicted nucleus bounding boxes (see Section \ref{promt:nuclei}), we segment the corresponding cells by employing the nuclei masks $\{y\}_{i=1}^m$ and the corresponding cell marker channel $t$ of image $x$ (see Figure~\ref{overview}, Step 2). In case that multiple cell marker channels are provided, we perform the cell segmentation process independently per channel. Subsequently, the resulting segmentation maps for each cell are combined by the confidence-weighted average of the channel-specific weights which reinforces high-confidence regions. An overview of our approach can be found in Algorithm \ref{algo:cellseg}. First, we define an initial search region for the cell segmentation using the nuclei mask $y_i$ where $i$ denotes the ith nuclei mask. An initial segmentation is then performed by sampling foreground points within a region 1.25 times the size of the nucleus bounding box and using them as prompts to a pretrained segmentation model $f(x,p)$.

\subsubsection{Recursive-Prompting Guided by Cellular Priors}
The initial mask is refined over a fixed number of iterations (see Figure~\ref{overview}, Step 2.1). This iterative process is designed to incorporate morphological and topological priors characteristic of cell images, promoting biologically plausible segmentations. The process is initiated from and anchored to the nucleus, which supports cell integrity by guiding the mask to grow as a single, contiguous object. To incorporate the topological prior of instance separation, the centers of neighboring nuclei are used as background (repulsive) points. This provides spatial context that helps delineate boundaries between adjacent cells and mitigates mask merging. New foreground points are sampled according to multiple criteria:
                \begin{itemize}
                    \item[-] \textit{Nucleus Anchor Points:} Points sampled randomly within the nucleus mask. These reinforce the primary anchor, ensuring the growing segmentation remains tethered to the correct cell and its identity.
                    \item[-] \textit{Hotspot Points:}  Points sampled from high-confidence regions (high logits) just outside the current mask boundary. These points guide the mask's expansion into plausible new areas of the cytoplasm, promoting the capture of the complete cell while respecting the cell integrity prior.
                    \item[-] \textit{Stabilizing Points:} Points selected from regions where the current prediction diverges from the previous iteration. These help to stabilize boundary refinement by discouraging oscillations and ensuring smoother convergence across iterations.
                \end{itemize}

In each iteration, the prompt p is constructed from these foreground points, the background points derived from neighboring nuclei, and a mean logit mask from the previous two iterations, which serves as a stabilizing spatial prior. This composite prompt is supplied to the segmentation model $f(x, p)$, which progressively refines the segmentation towards a converged state.

\begin{algorithm}[ht]
\caption{Cell Segmentation}
\begin{algorithmic}[1]
\Procedure{segmentCell}{$t$, $y_i$}
    \State background, foreground $\gets$ sampleInitialPoints($y_i$, $t$)
    \State segmentation, logits\_t $\gets$ f(x, \{foreground, background\})
    \State logits\_t\_minus\_1 $\gets$ logits\_t 
    \For {iter 1 \ldots I}
        \State prompt\_mask $\gets$ (logits\_t + logits\_t\_minus\_1) / 2
        \State foreground $\gets$ sampleGuidedForegroundPoints($y_i$, segmentation)
        \State background $\gets$ sampleBackgroundFromNeighbors($y_i$)
        \State p $\gets$ \{foreground, background, prompt\_mask\}
        \State logits\_t\_minus\_1 $\gets$ logits\_t 
        \State segmentation, logits\_t $\gets$ f(x, p)
    \EndFor
    \State \textbf{return} segmentation
\EndProcedure
\end{algorithmic}
\label{algo:cellseg}
\end{algorithm}

\subsubsection{Cell Mask Determination in Cell Population Context}
Iterative prompting per cell (Step 2.1) yields a binary mask per iteration (e.g., 7 iterations produce 7 binary masks for that cell). These masks are aggregated into a coverage map, where each pixel value reflects how often it was included across the iterations for a given cell. These per-cell coverage maps are then integrated into a global instance segmentation map (Figure~\ref{overview}, Step 2.2). Overlapping pixels are assigned to the cell with the highest coverage value. Pixels with low coverage in any cell's map (e.g., appearing in fewer than three of the seven masks, as depicted in Figure~\ref{overview} 2.2a-b) are discarded and assigned to the background. Finally, cells touching the image borders are excluded. This practice in HCS analysis is to prevent skewed feature measurements that would result from analyzing incomplete cells.

\subsection{Subcellular Entity Segmentation}
\label{sec:subcellularseg}

To segment subcellular entities, we use the previously generated cell masks (Section~\ref{sec:cellularseg}) and the dedicated subcellular marker channel, $ssm$. For each cell, its mask is used to crop the $ssm$ channel, isolating the search area. We then apply the automatic mask generation function of FastSAM to this cropped image to detect internal structures. The resulting masks are then re-projected to the full image's coordinates (see Appendix Algorithm~\ref{alg:subcell}).

    \begin{table}[h!]
    \centering
    \caption{Overview of the image datasets to evaluate the performance of subCellSAM.}
    \label{table:comparison}
    \begin{tabular}{l@{}ccc}
        \toprule
        \textbf{Dataset} & \textbf{\# Images} & \textbf{Biology} & \textbf{Ground truth} \\ 
        \midrule
        \multicolumn{3}{c}{\textbf{Datasets for cell segmentation evaluation}} & \textbf{Masks} \\
        \midrule
        BBBC008 \cite{ljosa2012annotated} & 12 & Human HT29 colon-cancer cells & CellProfiler \\ 
        \makecell[l]{Synthetic \\ Benchmark \cite{tang2024generative}} & 1502 & \makecell[c]{TREX/NXF1-Mediated\\ RNA Localization in the Nucleus \cite{zuckerman2020gene}} & \makecell[c]{Synthetically \\ generated}\\ 
        BBBC020 \cite{ljosa2012annotated} & 25 & \makecell[c]{Murine bone-marrow derived \\ macrophages} & CellProfiler \\
        \midrule
        \multicolumn{3}{c}{\textbf{Datasets for hit validation in drug discovery}}  & \textbf{Drug Potency}\\
        \midrule
        BBBC013 \cite{ljosa2012annotated} & 96 & \makecell[c]{Human U2OS cells \\ cytoplasm–nucleus translocation} & \makecell[c]{CellProfiler,\\ CNN} \\ 
        BBBC016 \cite{ljosa2012annotated} & 72 & \makecell[c]{Transfluor assay with GFP-tagged  \\ $\beta$-arrestin to track GPCR activity} & \makecell[c]{CellProfiler,\\ CNN} \\ 
        \bottomrule
    \end{tabular}
    \end{table}
    
\section{Experiments} 
    We conducted two different experiments: (1) evaluating the cell segmentation performance of subCellSAM, and (2) calculating downstream results for drug candidate validation in early drug discovery. To evaluate robustness, the parameters for subCellSAM were held constant across all experiments. Further performance improvements could likely be achieved through dataset-specific tuning. The parameters used are detailed in the Appendix. Each task required a distinct dataset due to their differing objectives and data requirements.

\subsection{Cell segmentation}
    
\subsubsection{Datasets}
    \label{sec:exp}
    Table \ref{table:comparison} shows an overview of all datasets used in the study. The first three lines in the table denote the datasets used for the analysis of cell segmentation performance. These datasets have a ground truth segmentation mask as provided by producers of the data, but lack information for hit validation, therefore we included a second set of data (see \ref{data_hit_main}). Further information about the biology and imaging procedures are in the Appendix (Section \ref{data_seg}). The 

    \subsubsection{Baseline Methods}
    We compare subCellSAM against a diverse set of baseline methods specifically developed for cell segmentation tasks: (1) CellPose 3 \cite{stringer2025cellpose3}, (2) DeepCell \cite{van2016deep}, (3) CellSAM \cite{israel2023foundationmodelcellsegmentation} and (4) CellProfiler \cite{Stirling2021}. Notably, the first three methods are trained on cellular images, whereas subCellSAM is applied directly without any fine-tuning or additional training. For more information, see Section~\ref{base_seg} in the Appendix.
   
    \subsubsection{Evaluation Metrics}
    To assess the performance and usability of subCellSAM for cell segmentation we employ two different metrics: the Dice Score (DSC) and the Intersection over Union (IoU). For more information about both metrics please see section \ref{eval_metr_seg} in the Appendix.

    \subsection{Hit validation in drug discovery}
 In drug discovery, hit validation verifies the biological activity of initial "hits" and assesses their suitability as drug candidates. High-content screening (HCS) supports this process through automated imaging: segmentation detects structures such as cells, and feature extraction quantifies attributes like shape, texture, and intensity. These features are essential for identifying cellular responses to treatment. Quantitative metrics such as the Z'-factor and EC\textsubscript{50}, derived from these features, are used to evaluate assay quality and compound potency (see Section~\ref{eval_metr_hit}) \cite{Barcelos2022, Nichols2006}.
    
    \subsubsection{Feature Generation}
    \label{feature}
    Following segmentation, morphological and intensity-based features are extracted from the binary masks at three hierarchical levels: cell, nucleus, and subcellular entity. For further information please see Section \ref{feat_gen} in the Appendix.

    \subsubsection{Datasets}
    \label{data_hit_main}
    The bottom part of Table \ref{table:comparison} denotes the datasets for the analysis of hit validation in drug discovery. These datasets provide a ground truth where known hit compounds are evaluated in a titration series to validate them. Biological information and imaging procedures are in the Appendix (Section \ref{data_hit}).

    \subsubsection{Baseline Methods}
    \label{hv:baseline}
    We evaluate the downstream performance of subCellSAM in comparison to a diverse set of established baseline methods commonly employed in hit validation assay analysis: (1) Genedata Imagence~\cite{STEIGELE2020812}, (2) CellProfiler~\cite{carpenter2006cellprofiler,logan2010screening} and (3) Multiscale CNN~\cite{godinez2017multi}. Details can be found in Section \ref{base_hit} of the Appendix.
    
    \subsubsection{Evaluation Metrics}
    To assess the performance and usability of subCellSAM for Downstream analysis we employ two different metrics: Z'-factor and EC\textsubscript{50} which are the predominant result read-outs in the biopharma industry for hit validation use cases. Please see section \ref{eval_metr_hit} in the Appendix for further detail.

    \section{Results and Discussion}
    We evaluate subCellSAM's performance on diverse datasets, focusing on cell segmentation and hit validation for early drug discovery. Comparisons with baseline models demonstrate its effectiveness and downstream impact.
    
    \subsection{Cell segmentation} 

    \begin{figure}[h!] 
    	\centering
        \includegraphics[width=1\linewidth]{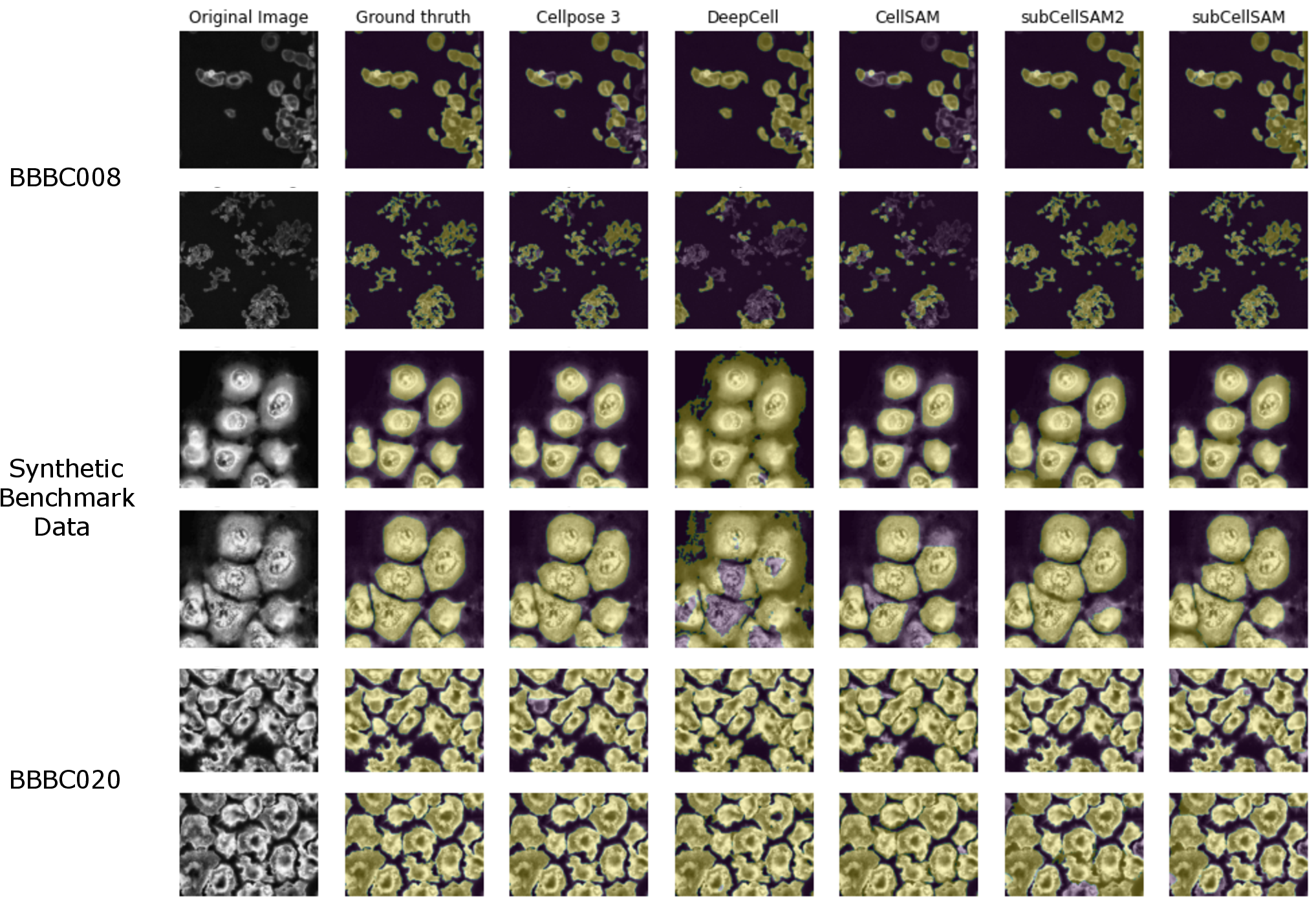}
        \caption{Overview of cell segmentation performance across three datasets. The first two columns show the original microscopy images (grayscale) and the corresponding ground truth masks (yellow). The remaining columns display segmentation masks from various methods (yellow), overlaid on the original images. Cellpose 3 (column 3), subCellSAM, and subCellSAM2 (columns 6 and 7) demonstrate consistently strong performance across all datasets. In contrast, DeepCell (column 4) and CellSAM (column 5) exhibit notable undersegmentation, particularly in the BBBC008 dataset (first two rows).}
        \label{fig:segmentation}
    \end{figure}

    \begin{table}[h!]
    \centering
    \caption{Comparison of the segmentation performance achieved by the different methods. Higher is better. Please note that mean DSC and IoU are computed over the entire mask, as the ground truth for BBBC008 and the Synthetic Benchmark contains non-separable cell masks, which precludes a per-instance metric calculation.}
    \label{table:comparison_transposed}
    \begin{tabular}{l@{}cccccc}
        \toprule
        \textbf{Method} & \multicolumn{2}{c}{\textbf{BBBC008} \cite{ljosa2012annotated}} & \multicolumn{2}{c}{\textbf{Synthetic Benchmark} \cite{tang2024generative}} & \multicolumn{2}{c}{\textbf{BBBC020} \cite{ljosa2012annotated}} \\ 
        \cmidrule(lr){2-3} \cmidrule(lr){4-5} \cmidrule(lr){6-7}
        & \makecell[c]{\textbf{mean}\\\textbf{DSC}} & \makecell[c]{\textbf{mean}\\\textbf{IoU}} & \makecell[c]{\textbf{mean}\\\textbf{DSC}} & \makecell[c]{\textbf{mean}\\\textbf{IoU}} & \makecell[c]{\textbf{mean}\\\textbf{DSC}} & \makecell[c]{\textbf{mean}\\\textbf{IoU}} \\
        \midrule
        CellPose 3 \cite{stringer2025cellpose3} & 0.887 & 0.801 & 0.899 & 0.827 & \textbf{0.898} & \textbf{0.819} \\ 
        DeepCell \cite{van2016deep} & 0.737 & 0.618 & 0.635 & 0.470 &  0.812 & 0.690 \\ 
        CellSAM \cite{israel2023foundationmodelcellsegmentation} & 0.602 & 0.484 & 0.837 & 0.739 & 0.817 & 0.694 \\ 
        \makecell[l]{CellProfiler \\(from \cite{tang2024generative})} & - & - & \textbf{0.922} & \textbf{0.856} & - & - \\ 
        \midrule
        \textbf{subCellSAM (ours)} & \textbf{0.901} & \textbf{0.832} & 0.892 & 0.807 & 0.868 & 0.772 \\ 
        \textbf{subCellSAM2 (ours)} & 0.900 & 0.821 & 0.828 & 0.714 & 0.851 & 0.745 \\ 
        \bottomrule
    \end{tabular}
    \end{table}
    
    Table~\ref{table:comparison_transposed} presents the mean Dice Score (DSC) and mean Intersection over Union (IoU) for subCellSAM and all baseline models across three datasets. Overall, subCellSAM demonstrates superior segmentation performance, outperforming all baselines on the BBBC008 dataset by 1.5\% in DSC and 3.9\% in IoU. On the Synthetic and BBBC020 datasets, subCellSAM maintains competitive performance, consistently surpassing both DeepCell and CellSAM. Interestingly, CellProfiler achieves the best results on the Synthetic dataset, likely due to the use of CellProfiler-generated masks as training input for StyleGAN2~\cite{Karras_2020_CVPR,zuckerman2020gene}.\\
    Figure~\ref{fig:segmentation} illustrates two example images per dataset alongside segmentation outputs from all evaluated methods. Notably, DeepCell and CellSAM tend to under-segment, particularly on the Synthetic and BBBC008 datasets (Figure~\ref{fig:segmentation}).\\
    It is important to note that all deep learning-based methods (i.e., all except CellProfiler) in Table~\ref{table:comparison_transposed} and Figure~\ref{fig:segmentation} were trained or fine-tuned using cell segmentation data. In contrast, subCellSAM operates in a zero-shot setting, relying on an in-context learning strategy that incorporates pre-defined morphological and topological priors through its prompting mechanism.\\
    The modular design of subCellSAM (see Section~\ref{sec:methods}) allows for flexible integration of different models. We evaluated both SAM-HQ~\cite{NEURIPS2023_5f828e38} and SAM2~\cite{ravi2025sam} for cell and subcellular segmentation, denoted as subCellSAM and subCellSAM2, respectively, in Table~\ref{table:comparison_transposed} and Figure~\ref{fig:segmentation}. Additional model details are provided in the Appendix (section \ref{mod_seg}).
    Interestingly, SAM-HQ consistently outperforms SAM2, despite the latter generally surpassing the original SAM model~\cite{ravi2025sam}. This may be attributed to SAM-HQ’s learnable HQ-Output token and global-local feature fusion mechanism~\cite{NEURIPS2023_5f828e38}, which enhance the segmentation of fine structures, an essential aspect of cell segmentation (see Figure~\ref{fig:segmentation}). Consequently, we use SAM-HQ for downstream hit validation analysis due to its slightly superior performance.\\
    Remarkably, the same parameter set was used across all datasets for subCellSAM segmentation (see Table~\ref{tab:workflow_para} in the Appendix for details). An ablation study on the BBBC008 dataset reveals that the parameters with the greatest impact are the ``number of prompts per cell'' and the ``percent coverage across prompts.'' A qualitative analysis of these parameters is presented in the Appendix (Figure~\ref{fig:ablation}). Fewer prompts result in smaller masks that may miss parts of the cell, while higher coverage thresholds yield more conservative masks by including only the most confidently predicted pixels.
    While dataset-specific tuning may improve performance, subCellSAM is designed for minimal adjustment, which is especially important given the frequent lack of ground truth data in real-world datasets.\\ For assays where a heterogeneous morphological cell response is expected, including variations in cell size, particularly in assays involving induced pluripotent stem cells (iPSCs) or primary cells, some parameter presets of subCellSAM may need to be adjusted, potentially on a per-cell basis. However, this is typically not necessary for hit validation assays.

    \begin{figure}[h!]
    	\centering
        \includegraphics[width=1\linewidth]{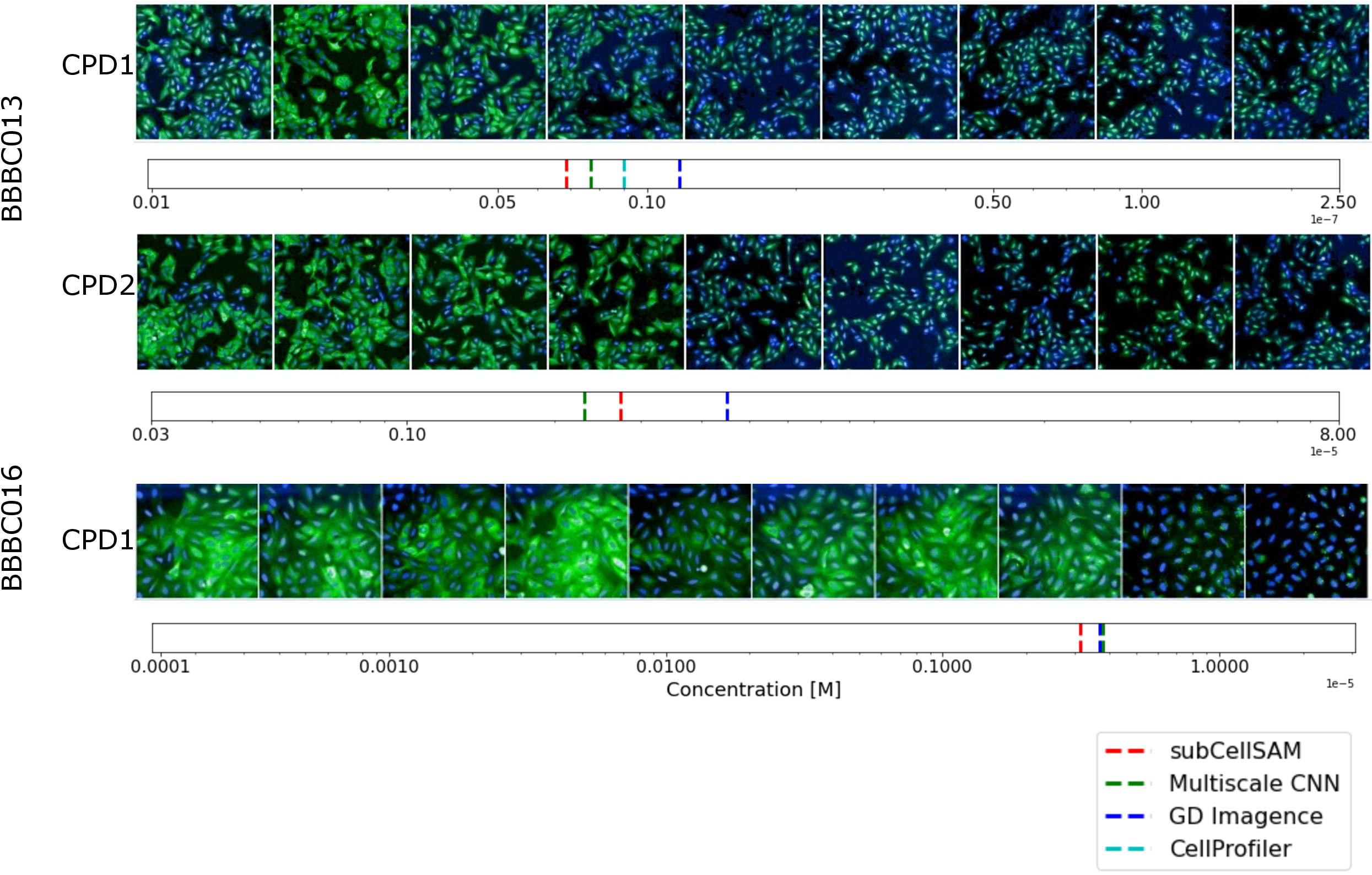}
        \caption{Overview of dosing behavior of the different compounds for both datasets with the calculated EC\textsubscript{50} values of the methods (see Appendix). The EC\textsubscript{50} values are shown in the bar under the respective images that show the nucleus (blue) and protein of interest (green). Note that not all values are available for all baseline methods (see Table~\ref{tab:sc_assay} in the Appendix).}
        \label{fig:ec50}
    \end{figure}

    \subsection{Application to hit validation in drug discovery} 

    Hit validation in drug discovery confirms that a compound truly affects the intended biological target and exhibits reproducible biological activity, rather than being a false positive~\cite{Barcelos2022}. To evaluate the performance of subCellSAM for hit validation, conventional image features were extracted from the generated masks of the cell, nucleus, and subcellular structures (see Section~\ref{feature}). The calculated metrics, the Z'-factor and EC\textsubscript{50} values, play a critical role in hit analysis and in preparing compounds for further filtering.\\
    Table~\ref{tab:sc_assay} in the Appendix presents both the maximum Z'-factor and the calculated EC\textsubscript{50} values for subCellSAM, compared to those obtained using baseline methods (see Section~\ref{hv:baseline}). The features extracted from the subCellSAM masks are on par with the baseline methods in terms of Z'-factor, indicating good assay quality and robustness.\\
    Notably, for the transfluor assay (BBBC016), subCellSAM demonstrates effective performance. Without any parameter tuning, subcellular structures are segmented effectively, resulting in improved signal-to-background ratios and, consequently, a high Z'-factor. The feature used in this case is the number of subcellular entities per cell (see Section \ref{feat_gen} in the Appendix).
    Regarding EC\textsubscript{50} values, the results obtained using subCellSAM are in agreement with all baseline methods (see Table~\ref{tab:sc_assay} in the Appendix). This is illustrated in Figure~\ref{fig:ec50}, which shows the EC\textsubscript{50} values from both baseline methods and subCellSAM, alongside representative images depicting cellular responses to compound dosing. All methods are in agreement and are able to detect the switch between start and endpoint cellular phenotypes (see Figure \ref{fig:ec50}). 

\section{Conclusion}

We introduce subCellSAM, a method for (sub)cellular segmentation in high-content screening that applies a foundation model in a zero-shot setting. The method's core is an in-context learning strategy that incorporates morphological and topological priors of cells to guide the segmentation process. Our experiments demonstrated that this approach yields segmentations that are competitive with specialized methods on three benchmark datasets. When applied to two industry-relevant hit validation tasks, the method produced high-quality downstream results without requiring dataset-specific parameter tuning. This suggests that leveraging structured, domain-specific priors within a general foundation model is a viable strategy for reducing manual configuration in automated HCS analysis pipelines, a conclusion supported by our use of a single hyperparameter set across all experiments.

\textbf{Limitations.} Our approach has two main limitations: First, the recursive prompting strategy leads to a significantly slower inference time compared to fine-tuned models. Second, subCellSAM requires a dedicated nucleus channel to initiate segmentation. Future work will aim to extend subCellSAM’s applicability from cell culture systems to more complex tissue samples.

\bibliographystyle{splncs04}
\bibliography{arxiv_subcell}

\newpage

\section{Appendix}

\subsection{Algorithm}
\begin{algorithm}
\caption{Nuclei Detection}\label{alg:euclid}
\begin{algorithmic}[1]
\Procedure{getNucleusMaskandCenter}{$nm$}
%\State candidateMasks $\gets$ predictMasks($x$)
\State candidateMasks $\gets$ sam($nm$)
\State nucleiMasks $\gets$ []
\State nucleiCenters $\gets$ []
\For {mask in candidateMasks}
	\State mask $\gets$ filterByShape(mask)
    \If{mask != null}
        \State nucleiCenter $\gets$ computeNucleiCenter(mask)
        \State nucleiMasks.add(mask), nucleiCenters.add(nucleiCenter)
    \EndIf	
\EndFor
\State \textbf{return} nucleiMasks, nucleiCenters
\EndProcedure
\end{algorithmic}
\end{algorithm}

\begin{algorithm}
\caption{Subcellular Segmentation}
\begin{algorithmic}[1]
\Procedure{segmentSubCell}{x, cell mask}
\State croppedImage $\gets$ crop($ssm$, cell mask)
\State segmentation $\gets$ g(croppedImage)
\State segmentation $\gets$ projectToOriginalImage(segmentation)
\State \textbf{return} segmentation
\EndProcedure
\end{algorithmic}
\label{alg:subcell}
\end{algorithm}

\subsection{Cell segmentation}

\subsubsection{Datasets}

\subsubsection{BBBC008}
\label{data_seg}
    The images stem from a study that developed a comprehensive lentiviral shRNA library targeting human and mouse genes, comprising 104,000 vectors for 22,000 genes \cite{Moffat2006}. This library enables efficient gene knockdown in various cell types, including non-dividing and primary cells. Applied to an arrayed viral high-content screen, it identified several known regulators and 100 candidate genes involved in mitotic progression in human cancer cells. Automated fluorescence microscopy and image analysis were used to detect mitotic cells and analyze cell images. The dataset includes 12 images of human HT29 colon-cancer cells with ground truth segmentation masks from CellProfiler \cite{ljosa2012annotated}. The samples were stained with Hoechst (channel 1), pH3 (channel 2), and phalloidin (channel 3). Hoechst stains DNA, highlighting the nucleus. Phalloidin stains actin, highlighting the cytoplasm. The pH3 stain indicates cells in mitosis. We use the phalloidin channel which stains the cell body for the benchmark. 

    \subsubsection{Synthetic Benchmark}
    The dataset introduces a novel benchmark framework applying generative adversarial networks (StyleGAN2 \cite{Karras_2020_CVPR} and Pix2PixHD \cite{Wang_2018_CVPR}) to create synthetic fluorescent cell images with controllable density and morphology, enabling systematic evaluation of cell segmentation methods \cite{tang2024generative}. It is trained on images stem from a study investigating the role of TREX and NXF1 export factors on the RNA distribution within the nucleus \cite{zuckerman2020gene}.  
    This approach provides a scalable, reproducible method for benchmarking segmentation algorithms in biomedical imaging. The ground truth masks were generated via the trained StyleGAN2. 
    
    \subsubsection{BBBC020}
    The images originate from a study investigating the role of Myd88 and MAPK in TLR-induced macrophage spreading, analyzed via automated imaging \cite{10.3389/fphys.2011.00071}. The study highlights the differential roles of Myd88 and MAPK pathways in early and late macrophage spreading responses. The image set includes 25 images, each with two channels. The samples were stained with DAPI for nuclei and CD11b/APC for the cell surface. We use the CD11b/APC stain for cell segmentation. The ground truth segmentation masks were produced using CellProfiler \cite{ljosa2012annotated}.

    \begin{table}[!ht]
        \centering
        \caption{Main workflow parameters applied for all data sets in this study}
        \label{tab:workflow_para}
        \begin{tabular}{l@{}cc}
            \toprule
            \textbf{Parameter} & \textbf{Value} \\ 
            \midrule
            \multicolumn{2}{c}{\textbf{Prompt loop}} \\
            num\_prompts\_per\_cell & 8 \\ 
            \multicolumn{2}{c}{\textbf{Points}} \\  
            max bbox\_area\_to\_sample & 1.5 \\ 
            num\_hotpoints & 4 \\ 
            \multicolumn{2}{c}{\textbf{Masks}} \\
            percentage\_coverage\_across\_prompts & 0.33 \\ 
            \bottomrule
        \end{tabular}
    \end{table}

\subsubsection{Models}
\label{mod_seg}
We use different models for the different parts in subCellSAM. For the initial nuclei segmentation, we use Fast Segment Anything \cite{zhao2023fastsegment}. For the cell- and subcellular segmentation we use either SAM-HQ \cite{NEURIPS2023_5f828e38} with ViT-h backbone or SAM2.1 \cite{ravi2025sam} with the Hiera-large backbone. 

\subsubsection{Baseline methods}
\label{base_seg}

\begin{enumerate} 
    \item CellPose 3 \cite{stringer2025cellpose3} utilizes a U-Net-style convolutional neural network with residual blocks following a restoration network to perform cell segmentation. It is widely regarded as one of the most advanced segmentation models, demonstrating state-of-the-art (SoTA) performance across a broad range of cell segmentation tasks and benchmarks \cite{Stringer2024.04.06.587952}.
    
    \item DeepCell \cite{van2016deep} employs a convolutional neural network (CNN) that classifies each pixel into one of three categories: background, boundary, or cell interior.
    
    \item CellSAM \cite{israel2023foundationmodelcellsegmentation} integrates a vision transformer backbone with a bounding box proposal model (DETR) and a SAM decoder for segmentation. The model was trained on approximately 1 million cell images.
    
    \item CellProfiler \cite{Stirling2021} relies on classical computer vision techniques—such as automatic thresholding and watershed algorithms—to construct cell segmentation pipelines. When well-parameterized, CellProfiler delivers strong performance and is often considered as gold standard \cite{tang2024generative}. 
\end{enumerate}

\subsubsection{Evaluation Metrics}
\label{eval_metr_seg}
\subsubsection{Dice Score}
    The Dice score is a statistical measure used to evaluate the similarity between two sets, particularly in the context of image segmentation:

    \[
    DSC = \frac{2|A \cap B|}{|A| + |B|}
    \]
    
    where \( |A \cap B| \) represents the number of common elements (pixels) between sets \( A \) (the segmentation result) and \( B \) (the ground truth mask), and \( |A| \) and \( |B| \) are the sizes of the sets. The Dice score ranges from 0 (no similarity) to 1 (perfect similarity). It is widely used to score object segmentation accuracy in biomedical imaging \cite{Bertels2019}.

\subsubsection{Intersection over Union (IoU)} is a metric used to assess the similarity between two sets, especially in image segmentation:

    \[
    \text{IoU} = \frac{|A \cap B|}{|A \cup B|}
    \]
    
    In this formula, \(|A \cap B|\) denotes the number of shared elements (pixels) between sets \(A\) (the segmentation output) and \(B\) (the ground truth mask), while \(|A \cup B|\) represents the total number of elements in both sets combined. The IoU score ranges from 0 (no overlap) to 1 (complete overlap). It is commonly employed to evaluate the accuracy of object segmentation in various imaging tasks. For the BBBC008 and Synthetic Benchmark datasets, the ground truth contains overlapping instances, which precludes a per-instance metric calculation. Therefore, for these datasets, the mean DSC and IoU are computed by comparing the entire predicted segmentation mask against the entire ground-truth mask.

\subsubsection{Ablation study}

\begin{figure}[!ht]
    	\centering
        \includegraphics[width=1\linewidth]{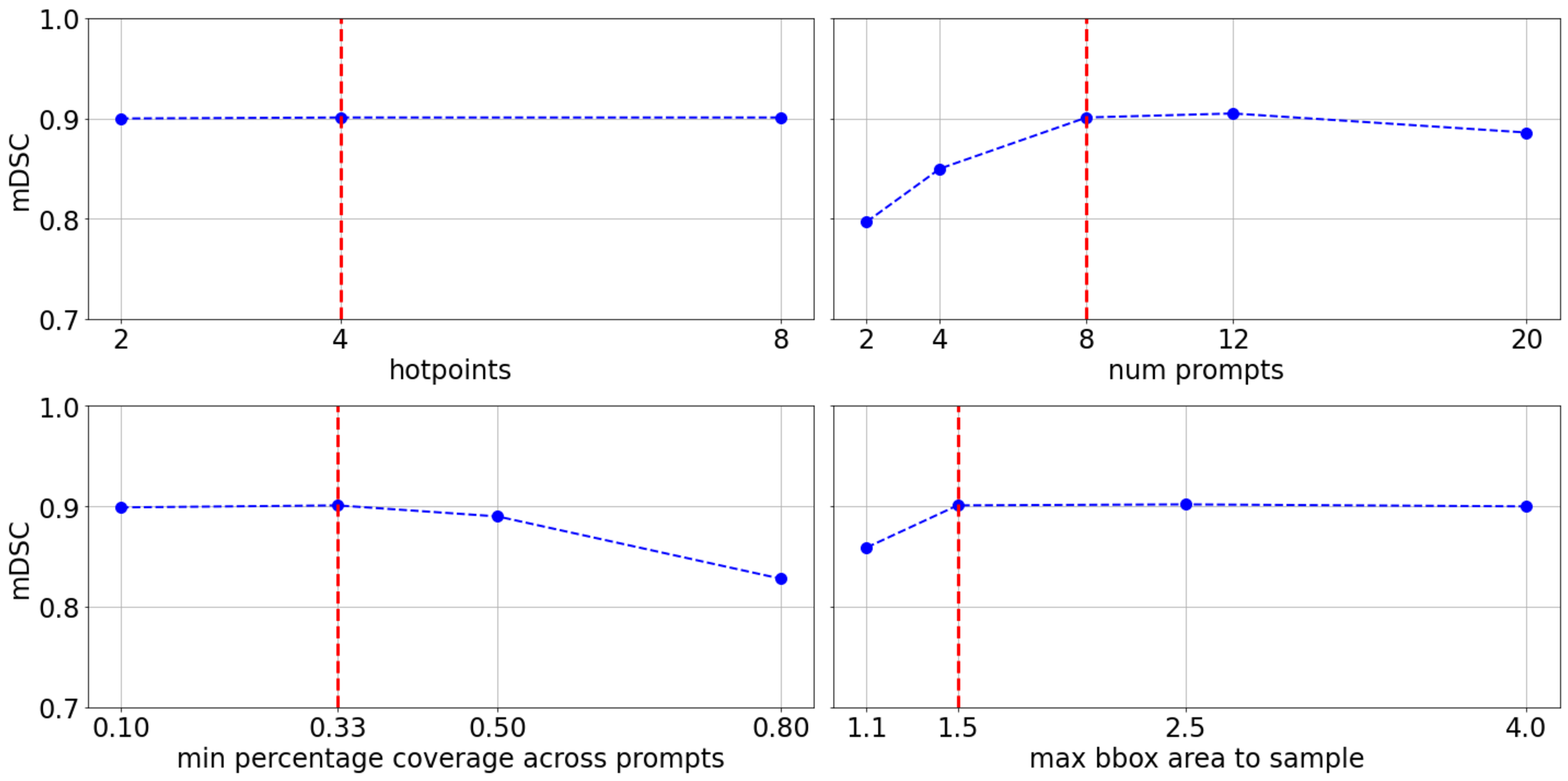}
        \caption{Overview of the influence of the main workflow parameters (see Table \ref{tab:workflow_para}) on the segmentation performance of subCellSAM on the BBBC008 dataset measured in mean DSC. Please note that the red vertical lines denote the setting used for the analysis of all datasets.}
        \label{fig:ablation}
\end{figure}

\begin{figure}[!ht]
    	\centering
        \includegraphics[width=1\linewidth]{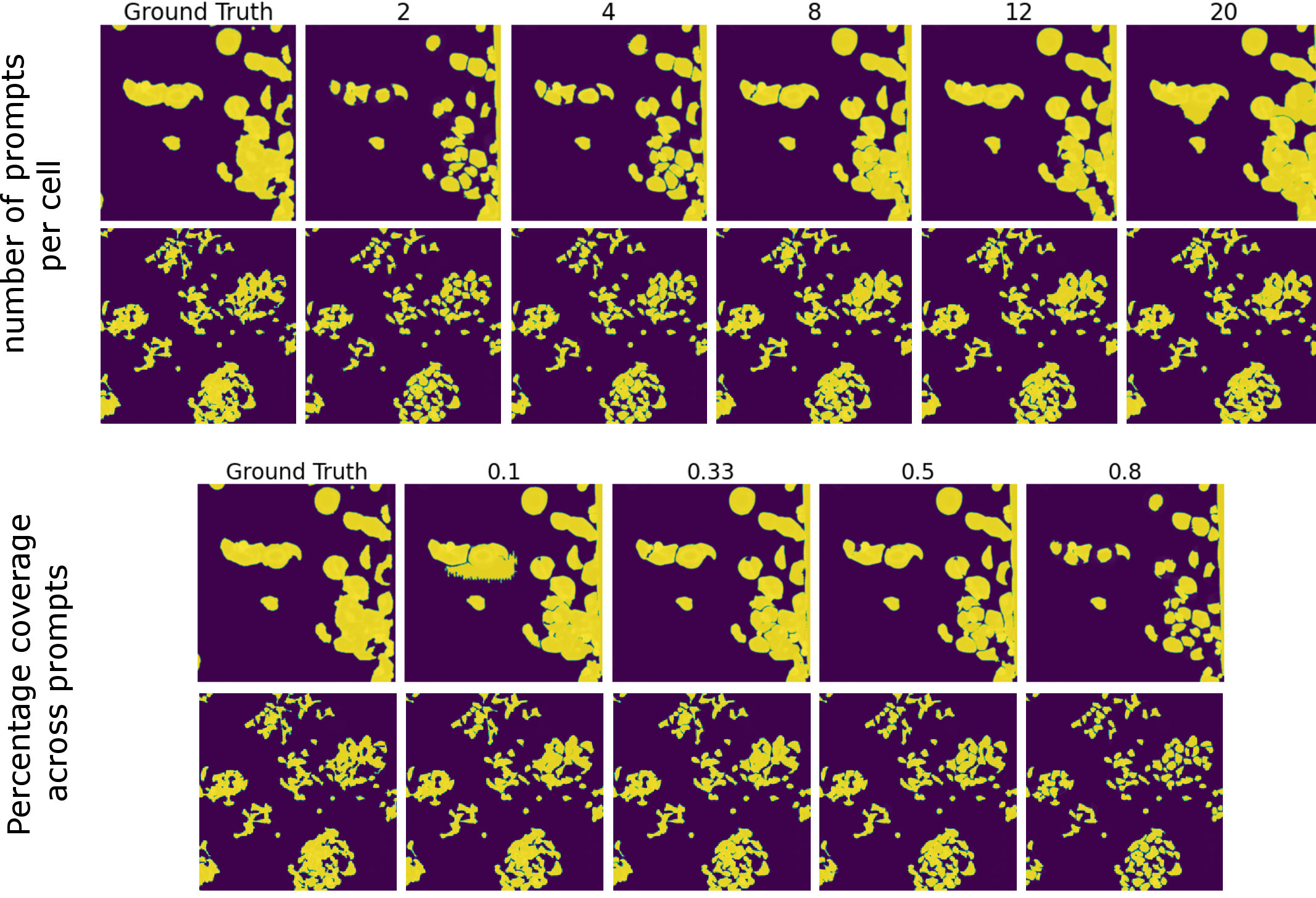}
        \caption{Example images of the segmentation results and the influenced by the most important parameters Number of prompts per cell and Percentage coverage across prompts.}
        \label{fig:ablation2}
\end{figure}

For the study of the influence of the main workflow parameters (Table \ref{tab:workflow_para}), we conducted an ablation study using BBBC008 segmentation performance as benchmark (see Figure \ref{fig:ablation} and \ref{fig:ablation2}). The most important parameters are the number of prompts that are carried out for each cell and the min percentage coverage across prompts that denotes the coverage needed to belong to a certain cell in post processing.

\subsection{Hit validation analysis}

\subsubsection{Datasets}
\label{data_hit}
\subsubsection{BBBC013}
    This dataset investigated the translocation of the Forkhead (FKHR-EGFP) fusion protein from the cytoplasm to the nucleus in stably transfected U2OS human osteosarcoma cells. In proliferating cells, FKHR is typically localized in the cytoplasm. However, even in the absence of external stimulation, FKHR continuously shuttles into the nucleus and is actively exported back to the cytoplasm by nuclear export proteins. When nuclear export is inhibited, FKHR accumulates in the nucleus. In this assay, nuclear export is blocked by inhibiting the PI3 kinase/PKB signaling pathway. Images were acquired with an IN Cell Analyzer 3000 with the Trafficking Data Analysis Module. One image was captured per channel: Channel 1 for FKHR-GFP and Channel 2 for DNA. Each image has a resolution of 640 × 640 pixels.
    
\subsubsection{BBBC016}
    The image data stems from an transflour assay which detects GPCR activation by tracking $\beta$-arrestin-GFP translocation in U2OS cells. Upon receptor activation, $\beta$-arrestin moves from the cytoplasm to endocytic vesicles, forming visible fluorescent puncta—an indicator of receptor internalization and signaling.
    The plate was analyzed using the Cellomics ArrayScan HCS Reader. Images were captured in 8-bit TIFF format, with separate channels for GFP (green) and DNA (blue). Each image has a resolution of 512 × 512 pixels.

\begin{table}[!ht]
       \centering
       \caption{Hit Validation Assay results}
       \label{tab:sc_assay}
       \begin{tabular}{l@{}ccccc}
           \toprule
           & \multicolumn{3}{c}{\textbf{BBBC013} \cite{ljosa2012annotated}}  & \multicolumn{2}{c}{\textbf{BBBC016} \cite{ljosa2012annotated}} \\
            \cmidrule(lr){2-4} \cmidrule(lr){5-6} \\
            \textbf{Method} & \textbf{Z'-factor $\uparrow$} & \makecell[c]{\textbf{EC\textsubscript{50}} \\ \textbf{CPD1 [M]}} & \makecell[c]{\textbf{EC\textsubscript{50}} \\ \textbf{CPD2 [M]}} & \textbf{Z'-factor $\uparrow$} & \makecell[c]{\textbf{EC\textsubscript{50}} \\ \textbf{CPD1 [M]}}\\
           \midrule
           Imagence (CNN) \cite{STEIGELE2020812}  & 0.774 & $1.16 \cdot 10^{-8}$ & $4.48 \cdot 10^{-6}$ & \textbf{1.00} & $3.78 \cdot 10^{-6}$  \\
           CellProfiler \cite{logan2010screening} & \textbf{0.920} & - & - & 0.430 & -  \\
           CellProfiler \cite{carpenter2006cellprofiler} & 0.910 & $9.00 \cdot 10^{-9}$ & - & - & -  \\
           Multiscale CNN \cite{godinez2017multi} & - & $7.70 \cdot 10^{-9}$ & $2.30 \cdot 10^{-6}$ & - & $3.80 \cdot 10^{-6}$ \\
           \midrule
           \textbf{subcellSAM (ours)} & 0.911 & $6.88 \cdot 10^{-9}$ & $2.73 \cdot 10^{-6}$ & 0.748 & $3.13 \cdot 10^{-6}$  \\
           \bottomrule
       \end{tabular}
\end{table}

\subsubsection{Baseline methods}
\label{base_hit}

\begin{enumerate}
    \item Genedata Imagence~\cite{STEIGELE2020812} leverages convolutional neural networks alongside an advanced training data generation pipeline to classify single-cell phenotypes. The resulting classification scores are subsequently used for calculating Z'-factors and EC\textsubscript{50} values.

    \item CellProfiler~\cite{carpenter2006cellprofiler,logan2010screening} applies traditional image analysis techniques to extract quantitative features from all segmented cellular compartments. These features serve as the basis for downstream analytical procedures.

    \item Multiscale CNN~\cite{godinez2017multi} processes input images at seven distinct resolutions in parallel, each through a dedicated three-layer convolutional neural network. The outputs are rescaled, concatenated, and passed through a final convolutional layer, followed by a fully connected layer with 512 units and a softmax layer to generate phenotype probability distributions. These probabilities are then utilized for the calculation of Z'-factors and EC\textsubscript{50} values.
\end{enumerate}
Please note that the segmentation baselines (CellPose 3, DeepCell, and CellSAM) cannot be used for Hit validation evaluation, as they lack intrinsic subcellular segmentation capabilities required for this task.

\subsubsection{Evaluation Metrics}
\label{eval_metr_hit}

\subsubsection{Z'-factor}

    The Z' factor \cite{zhang1999simple} is a metric used to assess the quality of screening assays, such as determining if the controls are appropriate for addressing specific biological questions. It is defined as:

    $$
    Z' = 1 - \frac{3 \cdot (\sigma_p + \sigma_n)}{|\mu_p + \mu_n|}
    $$
    
    where $\sigma_p$ and $\mu_p$ represent the sample standard deviation and sample mean of the positive (p) controls, respectively. The subscript $n$ refers to the neutral controls.A Z' factor value between 0.5 and 1 indicates an excellent assay, while values below 0.5 suggest the assay may need optimization.

    \subsubsection{EC50}

    The 50\% effective concentration (EC\textsubscript{50}) indicates the dosage of a drug required to achieve 50\% of its maximum biological effect. It is crucial to highlight that EC\textsubscript{50} is one of the most important and widely used metrics in screening assays for evaluating drug potency.

    The EC\textsubscript{50} is typically calculated by fitting the Hill equation to the observed data points:
    
    $$
    Y = S_0 + \frac{(S_{\infty} - S_0)}{1 + \left(\frac{\text{EC}_{50}}{[C]}\right)^n}
    $$
    
    where $S_0$ represents the fitted activity level at zero concentration of the test compound ("zero activity"), $S_{\infty}$ denotes the fitted activity level at infinite concentration of the test compound ("infinite activity"), $n$ is the Hill coefficient (indicating the slope at EC\textsubscript{50}), [C] is the concentration, and $Y$ is the activity.

\subsubsection{Feature generation}
\label{feat_gen}

In principle we extract the following features from the masks

\begin{itemize}
        \item \textbf{Nucleus Features:} Include size (area, perimeter, equivalent diameter), shape (e.g., eccentricity, solidity, extent, aspect ratio, circularity), orientation (major/minor axis lengths), and intensity from the nucleus channel.
        \item \textbf{Cell Features:} Include the same morphological descriptors as nuclei and intensity values from the cell marker channel.
        \item \textbf{Subcellular Entity Features:} Include size, shape, intensity statistics from the relevant channel, counts per cell, 
        \item \textbf{Correlation Features} the correlation of image signal of nucleus and cell marker channel is measured.
    \end{itemize}

Feature extraction was performed using implementations available in the scikit-image library \cite{van2014scikit}.

For the calculation of the Z'-factor and EC\textsubscript{50}, the following features are used after the mask calculation via subcellSAM and following feature extraction:

\begin{itemize}
  \item \textbf{BBBC013:} We use a Fisher Linear Discriminant Analysis (LDA) Weighted Average Feature composed of:
  \begin{itemize}
    \item Nucleus intensity
    \item Cell intensity
    \item Cell extent
    \item Cell perimeter
    \item Cell major axis length
    \item Cell minor axis length
    \item Nucleus protein correlation
  \end{itemize}
  The calculation is performed using Genedata Screener.
  
  \item \textbf{BBBC016:} We use the number of detected subcellular entities per cell.
\end{itemize}
These features possess the highest Z'-factor of all calculated features and were therefore chosen for the calculation of EC\textsubscript{50}.

\end{document}